\documentclass[useAMS,usenatbib]{mn2e}
\usepackage{graphicx}
\usepackage{epsf}
\usepackage{bm}
\usepackage{color}
\usepackage{mathrsfs}
\usepackage{lscape}
\usepackage{times}
\usepackage{hyperref}
\usepackage{amsmath}
\usepackage{longtable}
\usepackage{tabularx}
\usepackage{booktabs}
\usepackage{threeparttable}
\usepackage{array}
\usepackage{lscape}
\usepackage{amssymb}
\newcommand\aap{{ A}\&{A}}

\newcommand\aj{{AJ}}
\newcommand\apj{{ApJ}}
\newcommand\apjl{{\em ApJ}}
\newcommand\apjs{{ApJS}}

\newcommand\mnras{{MNRAS}}

\title[LSR from LSS-GAC] {Determination of the Local Standard of Rest using the LSS-GAC DR1}

\author[Y. Huang et al.]
               {Y. Huang$^{1}$\thanks{E-mails: yanghuang@pku.edu.cn (YH); x.liu@pku.edu.cn (XWL)},
                X.-W. Liu$^{1,2}$\footnotemark[1], H.-B. Yuan$^{2}$\thanks{LAMOST Fellow}, M.-S. Xiang$^{1}$, Z.-Y. Huo$^{3}$, B.-Q. Chen,
$^{2}$\footnotemark[2]
               \newauthor Y. Zhang$^{4}$, Y.-H. Hou$^{4}$\\
$^{1}$Department of Astronomy, Peking University, Beijing, 100871, P. R. China\\
$^{2}$Kavli Institute for Astronomy and Astrophysics, Peking University, Beijing, 100871, P. R. China\\
$^{3}$National Astronomical Observatories, Chinese Academy of Sciences, Beijing 100012, P. R. China\\
$^{4}$Nanjing Institute of Astronomical Optics \& Technology, National Astronomical Observatories, Chinese Academy of Sciences, Nanjing 210042, P. R. China}

\begin{document}

\date{}

\pagerange{\pageref{firstpage}--\pageref{lastpage}} \pubyear{2015}

\maketitle

\begin{abstract}
We re-estimate the peculiar velocity of the Sun  with respect to the local standard of rest (LSR) using a sample of local stars within 600\,pc of the Sun,  selected from the Large Sky Area Multi-Object Fiber Spectroscopic Telescope (LAMOST; also named the Guoshoujing Telescope) Spectroscopic Survey of the Galactic Anti-centre (LSS-GAC).
The sample consists of 94,332 FGK main-sequence stars with well-determined radial velocities and atmospheric parameters.
To derive the LSR, two independent analyses are applied to the data.
Firstly, we determine the solar motion by comparing the observed velocity distribution to that generated with the analytic formulism of  Sch\"onrich \& Binney that has been demonstrated to show excellent agreement with rigorous torus-based dynamics modelling by Binney \& McMillan.
Secondly, we propose that cold populations of thin disc stars, selected by applying an orbital eccentricity cut, can be directly used to determine the LSR without the need of asymmetric drift corrections.
Both approaches yield consistent results of solar motion in the direction of Galactic rotation, $V_{\odot}$, that are much higher than the standard value adopted hitherto, derived from Str\"{o}mgren's equation. 
The newly deduced values of $V_\odot$ are  1--2 km\,s$^{-1}$ smaller than the more recent estimates derived from the Geneva-Copenhagen Survey (GCS) sample of stars in the solar neighbourhood (within 100\,pc).
We attribute the small difference to the presence of several well-known moving groups in the GCS sample that, fortunately, hardly affect the LSS-GAC sample.
The newly derived radial ($U_{\odot}$) and vertical ($W_{\odot}$) components of the solar motion agree well with the previous studies.
In addition, for all components of the solar motion, the values yielded by stars of different spectral types in the LSS-GAC sample are consistent with each other, suggesting that the local disk is well relaxed and that the LSR reported in the current  work is robust.
 Our final recommended  LSR is,  $(U_{\odot},\,V_{\odot},\,W_{\odot})\,=\,(7.01\pm0.20,\,10.13\pm0.12,\,4.95\pm0.09)\,{\rm km\,s^{-1}}$.
\end{abstract}


\section{INTRODUCTION}

The local standard of rest (LSR) is defined a point in space that is moving on a perfectly circular orbit around the Galactic centre at the solar position.
An accurate estimate of the LSR, namely measuring the peculiar velocities of the Sun, is a starting point for Galactic kinematic and dynamic studies since all measurements are made relative to the Sun. 
It is needed for the derivations of basic parameters such as the Oort constants, which describe the local spatial variations of the stellar velocity field (Feast \& Whitelock 1997, Olling \& Dehnen 2003), as well as the orbital eccentricity distributions of different stellar populations which can be used to constrain the various scenarios of  Galactic disk formation (Sales et al. 2009, Wilson et al. 2011, Lee et al. 2011).
As a fundamental task of Galactic astronomy, an accurate determination of the three Galactic space velocity components of the solar motion with respect to the LSR, $U_{\odot}$, $V_{\odot}$ and $W_{\odot}$, has been a long-standing challenge. 
Similar to the ones given in Francis \& Anderson (2009) and Co\c{s}kuno\v{g}lu et al. (2011),  Table\,1 presents an updated summary of determinations (including results from the current work) of the solar motion from the literature of over a century.
As the Table 1 shows, in spite of utilizing a variety of methods and tracers, the measurements have hitherto failed to converge, especially for the component in the direction of Galactic rotation ($V_{\odot}$).

In principle, the radial ($U_{\odot}$) and vertical ($W_{\odot}$) components of the solar motion can be directly obtained by taking the negative of the mean heliocentric velocities of various stellar populations in the solar-neighbourhood  that are well relaxed.
The situation for the component in the direction of Galactic rotation ($V_{\odot}$) is more complicated, due to the non-negligible values of asymmetric drift, $V_{a}$, of stars with respect to the LSR -- motions that are related to the velocity dispersions of the individual stellar populations of concern.
The classical method to derive $V_{\odot}$ is based on Str\"omberg's equation (Str\"omberg 1946), which assumes a linear correlation between the mean negative heliocentric azimuthal velocity, $\overline{V_{\rm s}}=V_{\rm a}+{\rm V_{\odot}}$, and the squared radial velocity dispersion,  ${\sigma_{\rm R}^{2}}$, of any stellar sample.
Then a linear extrapolation to ${\sigma_{\rm R}^{2}}=0\,{\rm (km\,s^{-1})^{2}}$  for an idealized population yields  $\overline{V_{\rm s}} = V_{\odot}$.
Theoretically,  for any well-relaxed stellar population, the Jeans equation, i.e. Eq.\, (4.228) of Binney \& Tremaine (2008), links the asymmetric drift, $V_{a}$, with the radial velocity dispersion and the square bracket term (radial scale lengths and the properties of velocity ellipsoid) in axisymmetric equilibrium.
The physical underpinning of Str\"omberg's equation is that the square bracket term is a constant for the different stellar populations and the stellar populations used are in axisymmetric equilibrium. 
While the assumptions sound reasonable, the canonical value of $V_{\odot}=5.25\,{\rm km\,s^{-1}}$, derived by Dehnen and Binney (1998, hereafter DB98) based on  Str\"omberg's equation, has recently been found to be $\sim 6 - 7\,{\rm km\,s^{-1}}$ smaller than those determined using open clusters (Piskunov et al. 2006) and masers in massive star forming regions (Reid et al. 2009; Bobylev \& Bajkova 2010; McMillan \& Binney 2010) as tracers. 
The possible causes of discrepancies have been addressed recently by a number of investigators. 
Sch\"onrich et al. (2010) show that Str\"omberg relation could underestimate $V_\odot$ due to the presence of a  metallicity gradient, which breaks the linear relation between the mean rotation velocity and the squared velocity dispersion of stars binned by colour. 
This implies that the square bracket term may vary from one population to another population  as represented binned by the individual  colour bins.
In addition, the axisymmetric equilibrium of Galactic potential, a basic assumption underlying Str\"omberg's equation as mentioned above, can be broken due to the presence of  significant non-axisymmetric structures in our Galaxy (e.g. the spiral arms, the central bar; Francis \& Anderson 2009; Binney 2010).
Apart from  $V_{\odot}$, estimates of the radial and vertical components of the solar motion, $U_{\odot}$ and $W_{\odot}$, can also be biased by the presence of non-axisymmetric structures.

To minimize the effects of asymmetric drifts when estimating $V_{\odot}$, two different methods have been developed that avoid the assumption of a constant square bracket term for the different stellar populations underlying Str\"omberg's equation.
One is to determine the solar peculiar velocity from the offset of the modelled velocity distribution that best matches the observed data (Binney 2010; Sch\"onrich et al. 2010; Sch\"onrich \& Binney 2012, hereafter SB12). 
Alternatively, the LSR can also be determined utilizing cold populations of thin disc stars as tracers, under the assumption that they suffer from negligible asymmetric drifts (Francis \& Anderson 2009; Co\c skuno\u glu et al. 2011). 
Compared to the method based on Str\"{o}mberg's equation, the latter approach has  the advantage that it is simple and straight forward, and the results do not depend on the assumption that underpins the former method.  
The key to a successful application of this technique is to select and define a well representing sample of cold populations of thin disc stars.
It should be emphasised that all the approaches outlined above assume axisymmetric equilibrium.
Thus it is essential to select a sample of stars that are least affected by those known non-axisymmetric structures.
More discussions of this issue will be made below.

\begin{table*}
 \begin{center}
  \caption{Measurements of the LSR  in the literatures and from the current work}
  \begin{tabular}{cccccc}
  \hline
  Source                                                      & Data                                             &$U_{\odot}$         &$V_{\odot}$          &$W_{\odot}$      \\
                                                                      &                                                        & (km s$^{-1}$)       & (km s$^{-1}$)        & (km s$^{-1}$)     \\
   \hline 
   This study (2014)                                    &LSS-GAC DR1                                      &7.01$\pm$0.20   &10.13$\pm$0.12 &4.95$\pm$0.09  \\
   Bobylev \& Bajkova (2014)                & Young objects                            &6.00$\pm$0.50       & 10.60$\pm$0.80       & 6.50$\pm$0.30\\
   Co\c skuno\u glu et al. (2011)         & RAVE DR3                                    &8.50$\pm$0.29   &13.38$\pm$0.43   &6.49$\pm$0.26   \\
   Bobylev \& Bajkova (2010)                & Masers                                         &5.50$\pm$2.2     &11.00$\pm$1.70   & 8.50$\pm$1.20 \\
   Breddels et al. (2010)                           & RAVE DR2                                    &12.00$\pm$0.60     & 20.40$\pm$0.50     & 7.80$\pm$0.30      \\
   Sch\"onrich et al. (2010)                     & Hipparcos                                   &11.10$^{+0.69}_{-0.75}$ &12.24$^{+0.47}_{-0.47}$   &  7.25$^{+0.37}_{-0.36}$ \\
   Reid et al. (2009)                                   & Masers                                          &9.0                          &20                              &10                           \\
   Francis \& Anderson (2009)               & Hipparcos                                   &7.50$\pm$1.00        &  13.50$\pm$0.30      &  6.80$\pm$0.10       \\
   Bobylev \& Bajkova (2007)                & F \& G dwarfs                            & 8.70$\pm$0.50 &  6.20$\pm$2.22 &  7.20$\pm$0.80                     \\
   Piskunov et al. (2006)                           & Open clusters                             & 9.44$\pm$1.14 & 11.90$\pm$0.72 & 7.20$\pm$0.42       \\
   Mignard (2000)                                      & K0-K5                                             &       9.88 &      14.19 &       7.76 &                                                \\
   Dehnen \& Binney (1998)                   & Hipparcos                                    & 10.00 $\pm$0.36 & 5.25$\pm$0.62 & 7.17$\pm$0.38       \\
   Binney et al. (1997)& Stars near South Celestial Pole                              & 11.00$\pm$0.60 &  5.30$\pm$1.70 &  7.00$\pm$0.60                       \\
   Mihalas \& Binney (1981)& Galactic Astronomy (2nd Ed.)             &  9.00 &         12.00 &  7.0                               \\
   Homann (1886)& Solar neighborhood stars                                             & 17.40$\pm$11.2 & 16.90$\pm$10.90 & 3.60$\pm$2.30                 \\                                                                  
  \hline
\end{tabular}
\end{center}
\end{table*}

\begin{figure}
\centering
\includegraphics[scale=0.35,angle=0]{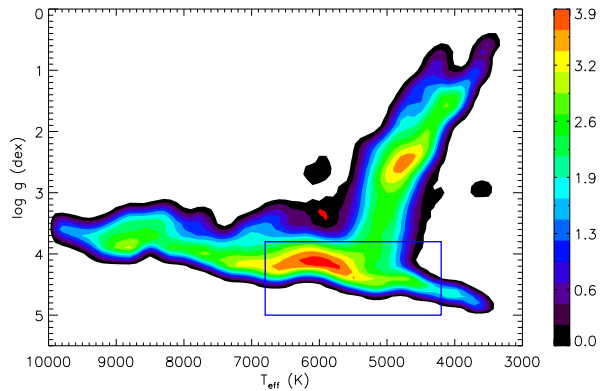}
\caption{Pseudo-color HR diagram of $\sim$ 0.7 million stars selected from the LSS-GAC DR1 that have spectral signa-to-noise ratios at 4650\,{\AA},  S/N(4650 \AA) $\ge 10$ and robust radial velocity and stellar parameter determinations. 
                   The colorbar shows the stellar density on a logarithmic scale. 
                   The blue box delineate the parameter space used to select FGK main-sequence stars analyzed in the current work.}
\end{figure}

\begin{figure*}
\centering
\includegraphics[scale=0.4,angle=0]{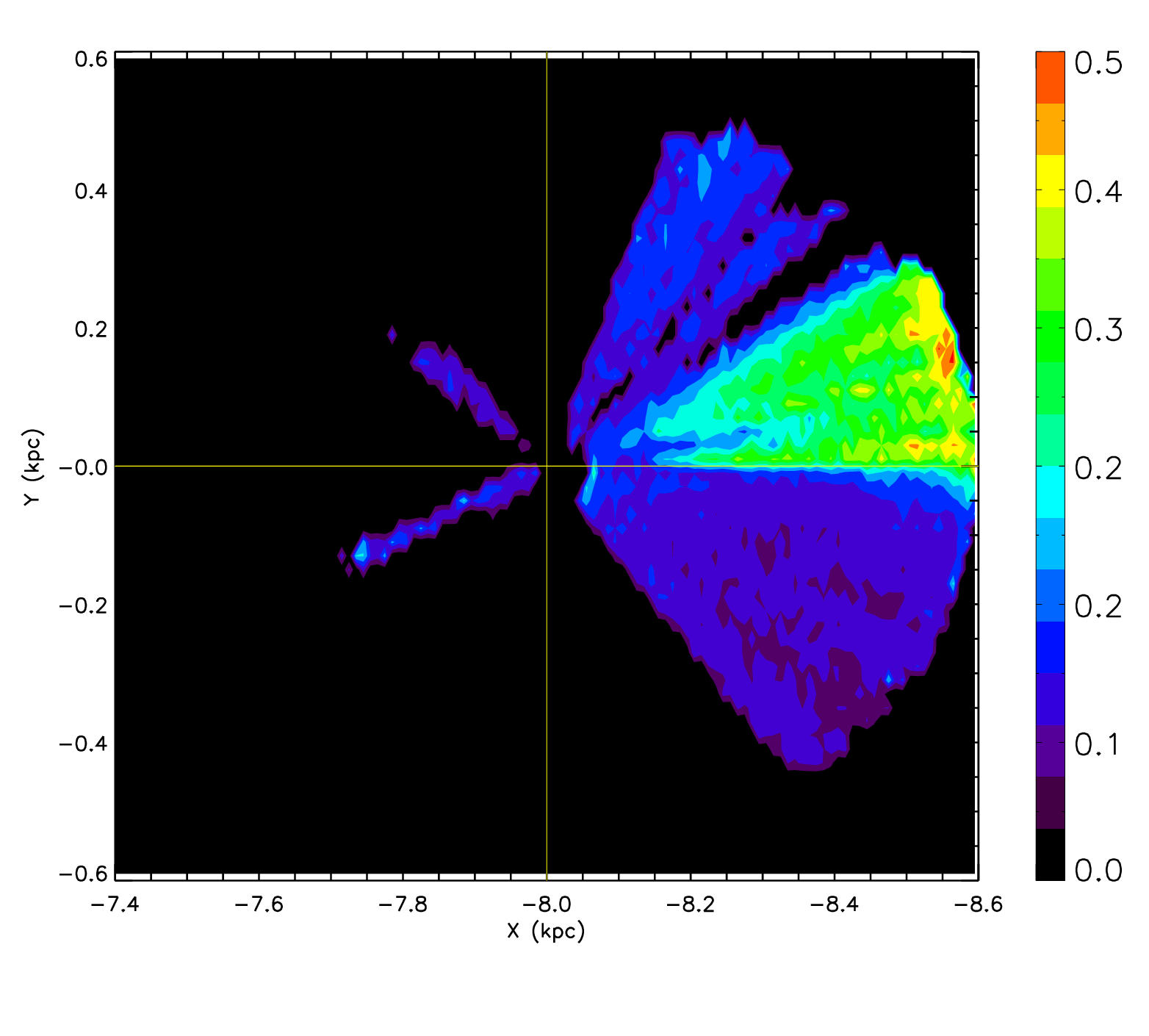}
\includegraphics[scale=0.4,angle=0]{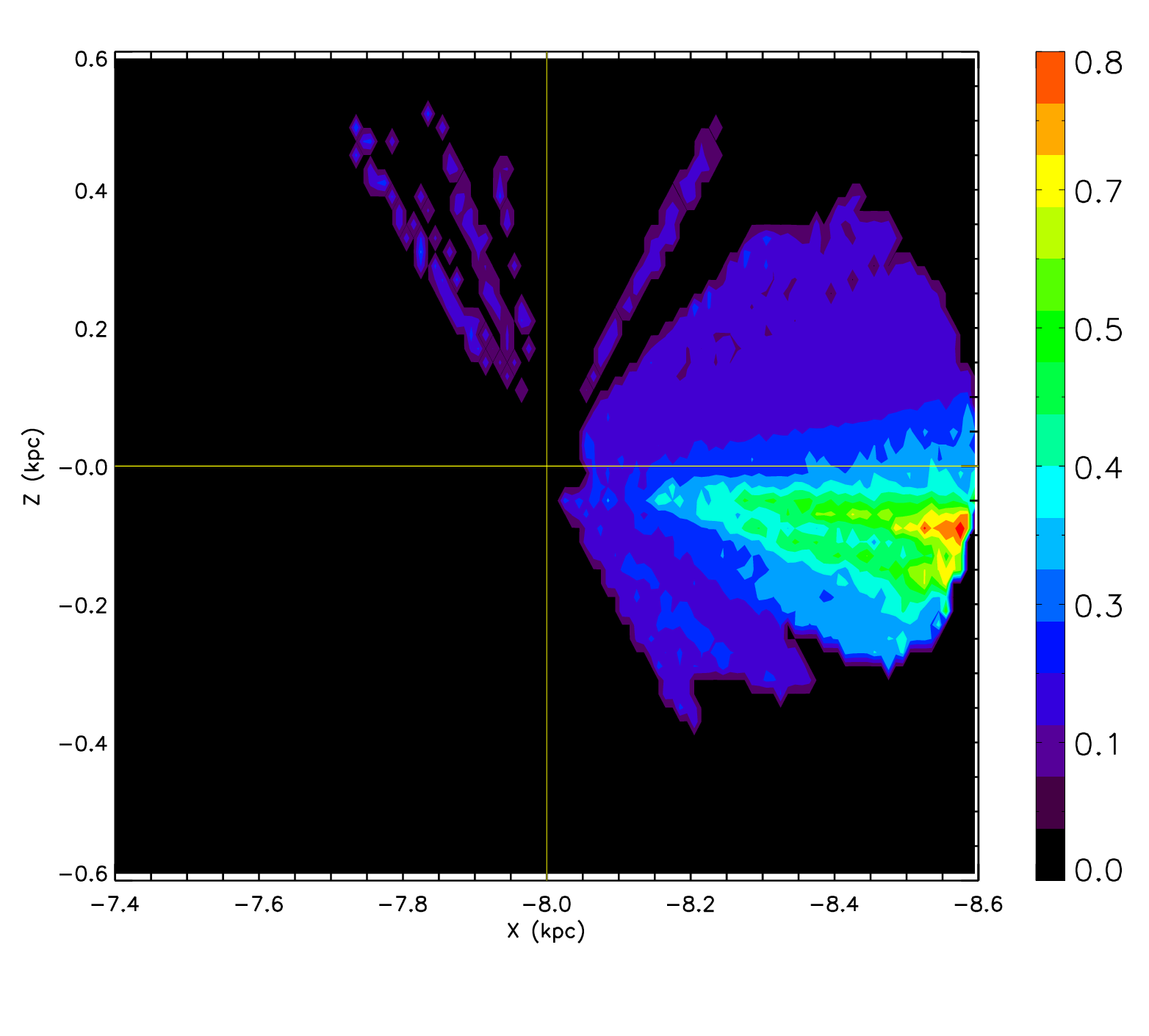}
\caption{Distributions of the integrated $E(B - V)$ reddening in the solar neighbourhood of 600\,pc for the current LSS-GAC DR1 sample of FGK stars in the Galactic $X$-$Y$ (left panel) and $X$-$Z$ (right panel) planes.
The colorbars indicate values of $E(B-V)$ in units of magnitude. }
\end{figure*}

In this paper, we re-estimate the peculiar velocity of the Sun with respect to the LSR by applying both the aforementioned methods to a sample of stars selected from the Large Sky Area Multi-Object Fiber Spectroscopic Telescope\footnote{LAMOST is a 4 meter quasi-meridian reflecting Schmidt telescope equipped with 4000 fibers, each of an angular diameter of 3\farcs3 projected on the sky,  distributed in a circular field of view (FoV) of 5$^{\circ}$ in diameter (Cui et al. 2012).} (LAMOST; also named the Guoshoujing Telescope) Spectroscopic Survey of the Galactic Anti-center (LSS-GAC, Liu et al. 2014).
For the velocity distribution fitting (VDF) method, we use the analytic formulism of SB12, which has been shown to reproduce the results of rigorous torus-based dynamics modelling (Binney \& McMillan 2011) to high fidelity.
As for the method based on cold populations of thin disc stars (CTDS), we use the orbital eccentricity, in replacement of the eccentricity vector adopted by Francis \& Anderson (2009), as a natural indicator of stellar population to select cold populations of  thin disc stars.
The sample selected from the LSS-GAC comprises FGK main-sequence stars that satisfy the following criteria:  (i) surface gravities $3.8\, <{\rm log}\,g<\,5.0$  dex; (ii) effective temperatures $4,200\,<T_{\rm eff}<\,6,800$ K; (iii) distances less than 600 pc; (iv)  total Galactic space velocity errors less than 15 km s$^{-1}$.
The first two constraints ensure that the stellar parameters of selected stars are well determined with the current LAMOST stellar parameter pipeline (cf. Section 3).
The third constraint is to minimize the effects of  large-scale streaming motions, induced by, for example, non-axisymmetric structures.
The fourth constraint is to ensure robust Galactic space velocities (cf. Section 4).  
To apply the CTDS method, we apply an additional cut to the sample: orbital eccentricities less than 0.13.
The constraint is pivotal to the method (cf. Section 6). 
Stars of very low orbital eccentricities move in near-circular orbits with negligible asymmetric drifts and thus can be used to determine $V_{\odot}$ without the need of correcting for their effects. 

The paper is organized as following. Section 2 presents a brief description of the LSS-GAC. 
The data are presented in Section 3. Section 4 describes the Galactic space velocities of sample stars and uncertainty control. 
Section 5 and 6 presents respectively the VDF and CTDS methods and their results.
The results are discussed in Section 7. 
Conclusions are given in the Section 8.

\section{LSS-GAC}
The LSS-GAC is a major component of the on-going LAMOST Galactic surveys, aiming to collect under dark and grey lunar conditions optical ($\lambda\lambda$3800-9000), low resolution ($R \sim 1, 800$) spectra for a statistically complete sample of over a million stars of all colors and of magnitudes $14.0 \leq r < 17.8$\,mag (18.5\,mag for limited fields), in a continuous sky area of $\sim$ 3, 400 sq.deg., centred on the Galactic anticentre (GAC), covering Galactic longitudes $150\,<l <\,210^{\circ}$ and latitudes $|b|\,<30^{\circ}$. 
Over 1.5 million very bright stars, brighter than $\sim$14.0\,mag., selected with a similar target selection algorithm, will also be  observed, utilizing bright lunar conditions.
The survey will deliver spectral classifications, values of stellar radial velocity $V_{\rm r}$ and atmospheric parameters (effective temperature $T_{\rm eff}$, surface gravity log $g$, metallicity [Fe/H]) for about 3 million Galactic stars.
The survey will sample main-sequence stars out to a few kpc, and tens of kpc for giants.
Combined with distances, e.g. the spectrophotometric distances of Yuan et al. (2015) and the proper motions from other surveys, the LSS-GAC provides a huge data set that {allows} one to study the stellar populations, kinematics and chemistry of the Galactic thin/thick disks and their interface with the halo in multi phase space (three dimensional space position and velocity, and metallicity). 
The Pilot and Regular Surveys of LSS-GAC were initiated in September of 2011 and 2012, respectively. 
The Regular Survey is expected to last for 5 years.
More details about the survey, including the scientific motivations,  target selections and data reduction, can be found in Liu et al. (2014) and Yuan et al. (2015).

\begin{figure}
\centering
\includegraphics[scale=0.35,angle=0]{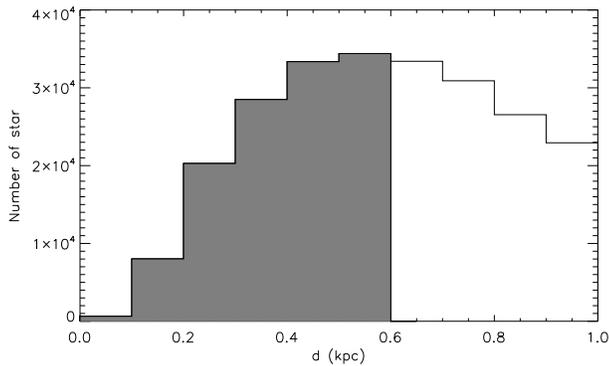}
\caption{Distribution of distances of FGK main-sequence stars in the LSS-GAC DR1 within 1\,kpc of the Sun. 
                   The shaded area shows stars within 600\,pc analyzed in the current work.}
\end{figure}

\section{DATA}
The data used in the current study are from the first data release of value-added catalogs of the LSS-GAC (hereafter DR1; Yuan et al. 2015).
There are about 0.7 million stars in the DR1 with a spectral signal-to-noise ratio per pixel ($\sim 1.07$\,\AA) at 4650\,{\AA}, ${\rm S/N\,(4650\,\AA)}\ge\,10$, and with robust determinations of stellar radial velocity $V_{\rm r}$ and atmospheric parameters ( $T_{\rm eff}$,  log $g$ and [Fe/H]) determined with the LAMOST Stellar Parameter Pipeline at Peking University (LSP3; Xiang et al. 2015a,b). 
The HR diagram of the stars is presented in Fig.\,1.
Parameters of FGK main-sequence stars are best determined with the LSP3.
Comparisons of results deduced from LAMOST multi-epoch duplicate observations, with results for common targets as given by the SDSS DR9 (Ahn et al. 2012), RAVE (Steinmetz et al. 2006) and APOGEE (Ahn et al. 2013) surveys and from the PASTEL compilation (Soubiran et al. 2010), as well as applying the Pipeline to open and globular clusters, show that for FGK dwarfs the LSP3 has achieved an accuracy of 5.0\,${\rm km\,s^{-1}}$, 150 K, 0.25 dex, 0.15 dex for $V_{\rm r}$,  $T_{\rm eff}$, log $g$ and [Fe/H], respectively.
We apply a surface gravity cut, $3.8\,<{\rm log}\,g<\,5.0$\,dex and an effective temperature cut, $4,200\,<T_{\rm eff}<\,6,800$\,K, to select the sample stars of FGK dwarfs. In total, $\sim$ 0.4 million stars are selected, as rounded by the blue box in Fig.\,1.
To test the robustness of the methods used in the current work, we further divide the total sample into three sub-samples, consisting respectively of stars of spectral type F, G and K, based on the effective temperatures of the stars: F-type of $6,000\le$\,$T_{\rm eff}$\,$\le 6,800$\,K, G-type of $5,300\le$\,$T_{\rm eff}$\,$< 6,000$\,K and K-type of $4,200\le$\,$T_{\rm eff}$\,$< 5,300$\,K and apply those methods to each sub-sample.

\begin{figure*}
\begin{center}
\includegraphics[scale=0.4,angle=0]{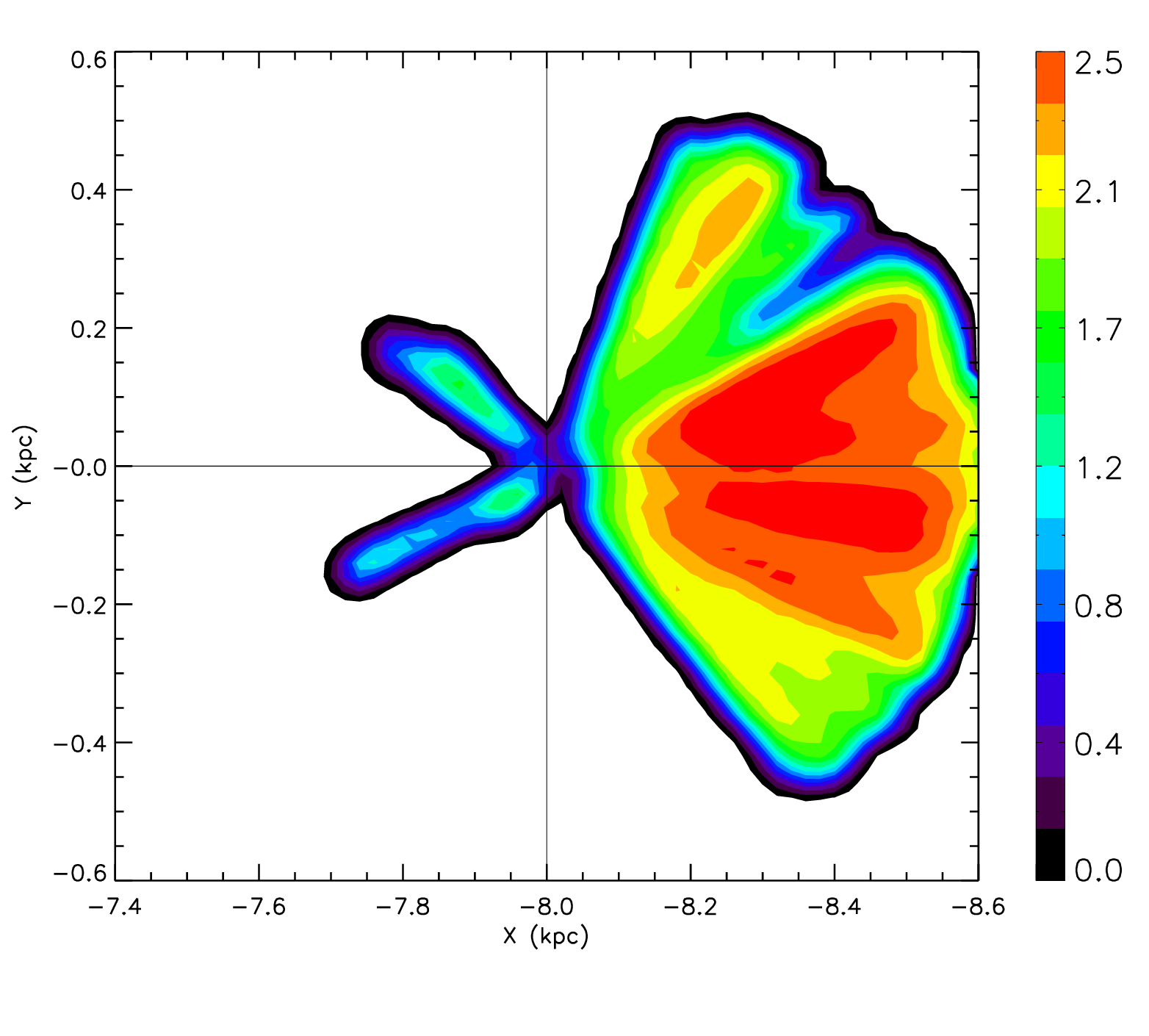}
\includegraphics[scale=0.4,angle=0]{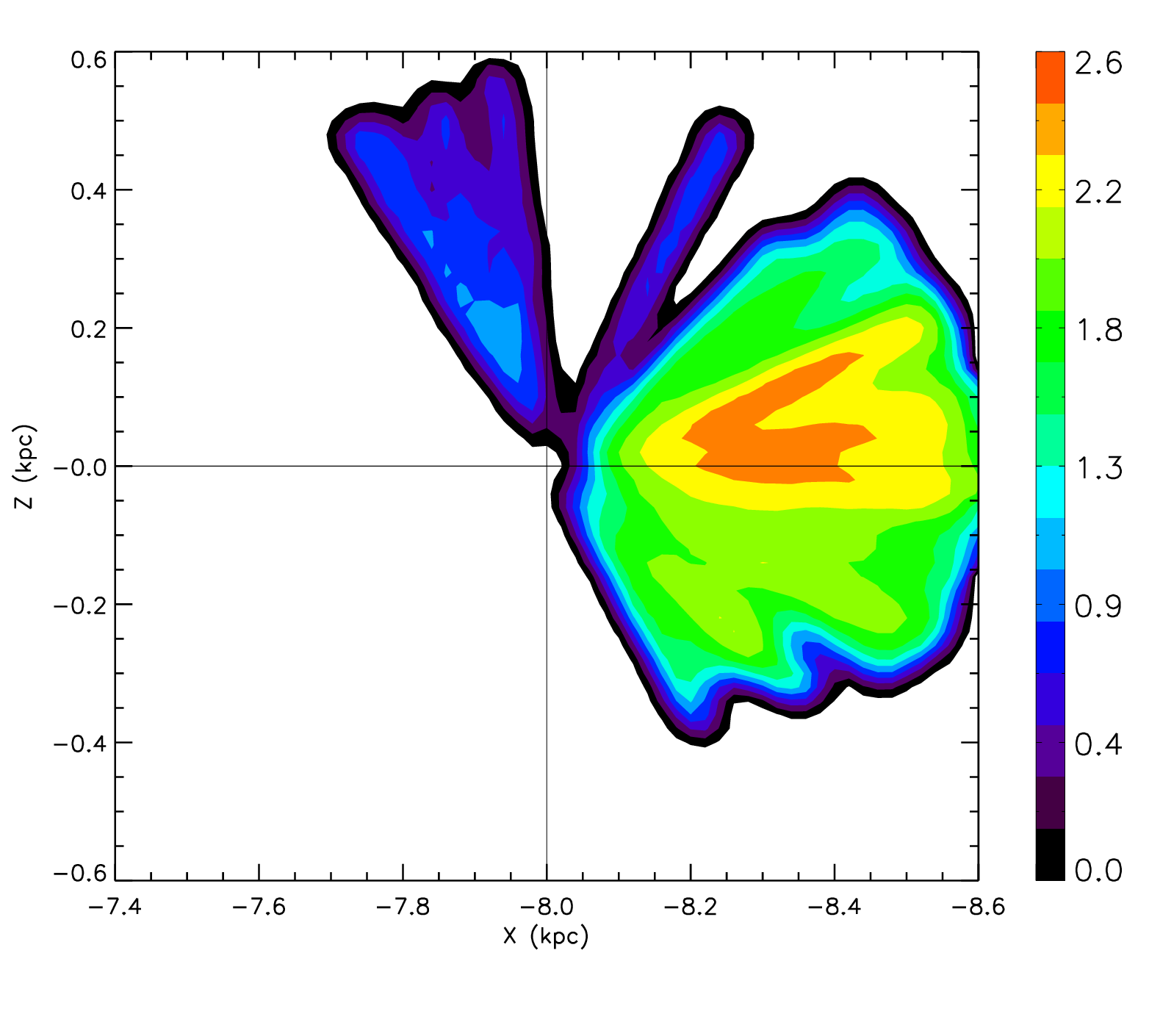}
\caption{Spatial distributions of the sample stars in the $X$ -- $Y$ (left panel) and $X$ -- $Z$ (right panel) planes.
                   The logarithmic stellar number densities in units of number per kpc$^{2}$ are indicated by the side colorbars.}
\end{center}
\end{figure*}

\subsection{Extinctions and distances}
Since most of the sample stars fall in the Galactic disc at distances beyond 100 pc (see Fig.\,3), corrections for the interstellar extinction by dust grains are essential for robust distance determinations.
Here we adopt estimates of extinction for individual stars as deduced from the photometric colours with the ``standard pair'' technique (Stecher 1965; Massa et al. 1983; Yuan et al. 2013, 2015).
A comparison with the values given by the extinction map of  Schlegel et al. (1998) for high Galactic latitude regions shows the technique has achieved a precision of about 0.04 mag in $E(B-V)$\footnote{For targets without optical photometry, the estimated uncertainties of extinction increase to about 0.1\,mag in $E(B-V)$ (Yuan et al. 2015).}.
Fig.\,2 shows the distributions of integrated $E(B - V)$ out to a distance of 600 pc in the $X$-$Y$ and $X$-$Z$ Galactic planes for the sample stars. 
As expected, the reddening increases with distance. The local bubble of very low extinction is also clearly visible in quadrants of $Y<\,0$ out to 600 pc.

For distances, geometric parallaxes are not available for the majority of stars targeted by the LSS-GAC. 
We have thus developed an empirical relation between the absolute magnitudes and the atmospheric parameters ($T_{\rm eff}$, log $g$ and [Fe/H]), based on the MILES (S\'anchez-Bl\'azquez et al. 2006) spectral library (Yuan et al. 2015).
The relation is based on the absolute magnitudes derived from the Hipparcos parallaxes (typical errors $\le$\,10 per cent) and the atmospheric parameters compiled and homogenized by Cenarro et al. (2007), mostly deduced from high spectral resolution measurements.
We find that distances thus derived for FGK dwarfs are better than 15 per cent and show no obvious systematic errors.
The distribution of distances of FGK dwarfs within 1 kpc in the DR1 is presented in Fig.\,3. 
For the purpose of current work, only stars within 600 pc, as shown by the shaded area are analyzed.

Fig.\,4 shows the spatial distributions of the selected stars in the Galactic planes for a right-handed Cartesian coordinate system centred on the Galactic centre, where the $X$-axis passes through the Sun and points towards the Galactic centre, the $Y$-axis is in the direction of Galactic rotation and the $Z$-axis points towards the North Galactic Pole (NGP).
The Sun is located at ($X$, $Y$, $Z$) = ($-R_{0}$, $0$, $0$), where $R_{0} = 8$\,kpc.
Most of the stars fall in the anti-centre direction, with the rest coming from the Very Bright (VB) plates observed under bright lunar conditions.

\subsection{Radial velocities and proper motions}
\vspace{-5.7pt}
As mentioned above, for FGK dwarfs radial velocities yielded by the LSP3 are accurate to $\sim$ 5 km s$^{-1}$.
However, we find that for common stars targeted by both surveys, the LSS-GAC velocities yielded by the LSP3 have  an offset of 3.1 km s$^{-1}$ relative to those of APOGEE, the latter have a measurement accuracy better than 0.15 km s$^{-1}$.
Hence, we have applied an offset correction to all radial velocities yielded by the LSP3 (cf. Xiang et al. 2015b for a detailed description). 

The proper motions of our sample stars are taken from the Fourth United States Naval Observatory (USNO) CCD Astrograph Catalog (UCAC4; Zacharias et al. 2013), an all-sky astrometric catalog archiving over 113 million objects. Among them, over 105 million stars have proper motions. The catalog is completed to $R\sim$ 16 mag. 
Typical random and systematic errors of the UCAC4 proper motions are 4 and 1-- 4 mas yr$^{-1}$,  respectively (Zacharias et al. 2013).

\begin{table*}
 \begin{center} 
  \caption{Coefficients of a two-dimensional second-order polynomials fit to the systematic errors of proper motions of UCAC4  as a function of the Right Ascension $\alpha$ and Declination $\delta$.}
  \begin{threeparttable}
  \begin{tabular}{lccccccc}
  \hline
   Components &                           $a_{0}$ & $a_{1}$      &$a_{2}$     &$a_{3}$       & $a_{4}$&$a_{5}$ \\
   \hline
   $\mu_{\alpha}\cos \delta$  &1.29  &-0.0137&$3.27\times10^{-5}$&$-4.90\times10^{-5}$&$-0.0106$&$5.01\times10^{-5}$\\ 
   $\mu_{\delta}$                            & $ -1.82$&$3.68\times10^{-3}$&$-2.91\times10^{-6}$&$-5.57\times10^{-5}$&$-4.70\times10^{-3}$&$2.35\times10^{-4}$\\        
 \hline
 \end{tabular}
\begin{tablenotes}
\item[]\textbf{Note}: The fitting function has the form: $\mu_{\rm fit}=a_{0}+a_{1}\alpha+a_{2}\alpha^{2}+a_{3}\alpha\delta+a_{4}\delta+a_{5}\delta^{2}$, where $\mu_{\rm fit}$, in mas yr$^{-1}$, represent the the two components of quasar proper motions, $\mu_\alpha\cos\delta$ and $\mu_\delta$, and $\alpha$ and $\delta$, both in units of degree, are the Right Ascension and Declination.
\end{tablenotes}
  \end{threeparttable}
  \end{center}
\end{table*}

For our sample, a systematic error of a few mas yr$^{-1}$  in proper motions could induce a bias of several km s$^{-1}$ in the determination of LSR, especially for the $V_{\odot}$ and $W_{\odot}$ components.
To correct for the systematic errors of UCAC4 proper motions,  we have cross-matched  the UCAC4 with the 2nd release of Large Quasar Astrometric Catalog\footnote{Quasars, being distant extragalactic point sources, have essentially zero proper motions and are thus the best objects to estimate and correct for uncertainties of stellar proper motion measurements.} (LQAC, Souchay et al. 2012).
After excluding quasars with poor proper motion determinations, this yields a total of 1,700 quasars distributed over the whole sky.
Corrections to account for the systematic errors of UCAC4 proper motions are then derived by fitting the proper motions of quasars (by default, quasars should have zero proper motions) with a two-dimensional second-order polynomials as a function of the Right Ascension and Declination\footnote{Similar corrections for the PPMXL (Roeser et al. 2010) proper motions have been obtained by Carlin et al. (2013).}.
Table 2 presents the fit coefficients.
No magnitude and color dependences of the systematic errors of proper motions of  UCAC4 are found for quasars of $R$ magnitudes between 12 and 16 mag and $B - V$ color between 0.1 and 1.4\,mag.
UCAC4 proper motions corrected for the systematic errors, as calculated from the above fits, are used for all calculations given below.

\begin{figure}
\begin{center}
\includegraphics[scale=0.55,angle=0]{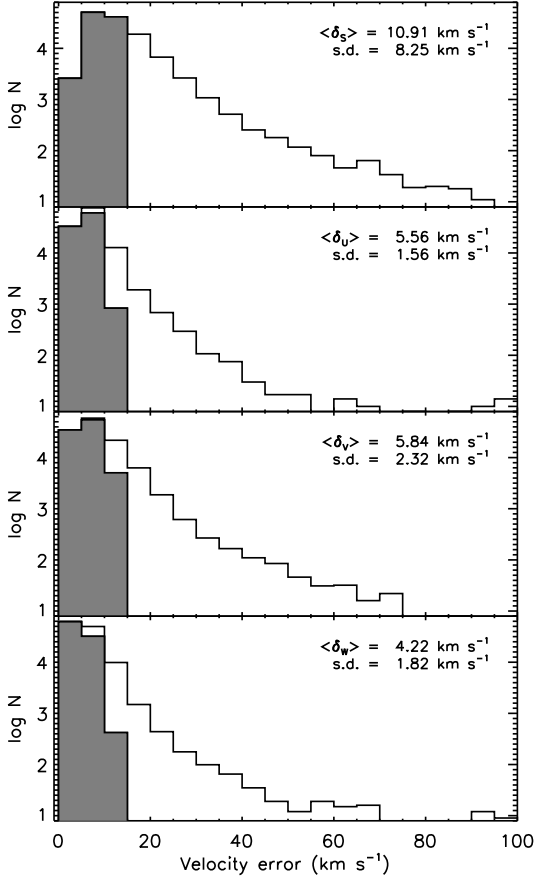}
\caption{Histograms of Galactic { space} velocity uncertainties. 
                  The top panel shows the total { Galactic space} velocities and the rest the three components.
                  Stars with a total { Galactic space} velocity error $\delta_{S}>\,15$ km s$^{-1}$ are removed from the sample. 
                  The histogram distributions of the remaining stars are represented by the shaded areas.
                  The mean and standard deviation of errors of the total velocities marked in the top panel are calculated before applying the cut, whereas those in the rest panels for the three velocity components are calculated  after applying the cut.}
\end{center}
\end{figure}

\section{ Galactic space velocity and uncertainty control}
{ From the distances calculated with the empirical relation described in Section 3., the LSP3 radial velocities and the UCAC4 proper motions, with the latter two corrected for the systematic errors as discussed in Section\,3.2,} we calculate the total Galactic space velocities and their three components, ($U$, $V$, $W$) of the sample stars using the standard transformation matrices derived by Johnson \& Soderblom (1987).
$U$, $V$ and $W$ refer to respectively the velocity components in the $X$, $Y$ and $Z$ directions in a right-handed Cartesian coordinate system as defined in Section 3.1.

Uncertainties of the three velocity components, $\delta_{U}$, $\delta_{V}$ and $\delta_{W}$, are also calculated using the standard algorithm given by Johnson \& Soderblom (1987) by propagating the uncertainties of the radial velocities, proper motions and distances.
The uncertainties of total { Galactic space} velocities are then calculated using,
\begin{equation}
\delta_{S}^{2} = \delta_{U}^{2} + \delta_{V}^{2} + \delta_{W}^{2},
\end{equation}

The distributions of uncertainties of the total { Galactic space} velocities and their three components are presented in Fig.\,5.
The mean and standard deviation of uncertainties of the total { Galactic space} velocities are {$\langle\delta_{S}\rangle  = 10.91$ km s$^{-1}$} and ${\rm s.d.} = 8.25$ km s$^{-1}$, respectively.
Stars with $\delta_{S}>\,15$ km s$^{-1}$ are discarded, totaling 30,829 stars, about 25\% of the sample.
After applying the cut, the average uncertainties of the three velocity components for the remaining stars of the sample are reduced to about 5 km s$^{-1}$, with a standard deviation of $\sim 2$\,km s$^{-1}$.
The velocities of the remaining sample stars are thus sufficient accurate to make a robust determination of the LSR.

\section{Determining $V_{\odot}$ by VDF}
\subsection{Bulk motion effects}
As pointed out earlier, determining the solar peculiar velocity in the Galactic rotation direction, $V_{\odot}$, is not as straightforward as for those in the other two directions ($U_{\odot}$ and $W_{\odot}$), due to the presence of significant asymmetric drift  in the azimuthal direction, $V_{\rm a}$.
The classical linear Str\"omberg's equation was proposed to solve this problem (e.g. DB98).
{However, as pointed out by Sch\"onrich et al. (2010), Str\"omberg's equation can be broken due to the presence of a metallicity gradient that breaks the linear relation between asymmetric drift and radial velocity dispersion of stars binned by colour.}
To overcome the obstacle, a technique has been developed to derive $V_\odot$ from the offset of { the velocity distribution} of { a dynamical model} that  best matches the measurements.
Using this technique, Binney (2010) constructs an azimuthal velocity distribution based on actions estimated from the adiabatic invariance of the local sample of stars of the Geneva-Copenhagen Survey (GCS, Nordstr\"om et al. 2004; Holmberg, Nordstr\"om \& Andersen 2009), and revises $V_{\odot}$ upwards to 11 km s$^{-1}$.
Similarly, by chemodynamically modeling the GCS sample, Sch\"onrich et al. (2010) conclude that { DB98 underestimated  $V_{\odot}$ by $\sim 7$\,km s$^{-1}$ .}
{ As described earlier, the VDF method is also founded on the assumption of axisymmetric equilibrium.
On the other hand, it has been known that non-axisymmetric structures (e.g. the spiral arms, the central bar) could distort the observed velocity distribution (Dehnen 1999; De Simone, Wu \& Tremaine 2004; Antoja et al. 2009; Minchev et al. 2009).
For the moment, the only way to minimize the dynamical effects of those non-axisymmetric structures is to carefully select a sample of stars least affected by those structures.}
Unfortunately, a number of moving groups have been found  in the ($U, V$) velocity space in the GCS sample (e.g. Dehnen, 1998), which is restricted to the immediate vicinity of the Sun (within 100\,pc) and consists of just over ten thousand stars of a variety of spectral types.
As shown in Fig. 6, the distribution of the $V$ velocity component of GCS stars is seriously distorted by four well-known moving groups, especially the Pleiades and/or Hyades streams.
Unless the effects of bulk motions of those streams could be properly modeled, the aforementioned determinations of $V_{\odot}$ based on the GCS sample could be significantly biased.
Unfortunately, the origin of those moving groups is still on debate (e.g.  Antoja et al. 2011), making it difficult to either incorporate them in the dynamical model or remove their effects by subtracting the overdensities produced by them in the observed velocity distribution.
Another large sample of local stars, alternative to the GCS one and less plagued by bulk motions of moving groups, is needed to apply this powerful technique developed by Biney (2010) in order to accurate determine $V_{\odot}$ without biases.
The LSS-GAC sample of local stars presented in the current work seems to exactly fill the bill. 
As shown in Fig. 7, the distribution of $V_\phi$ velocity {components} of our LSS-GAC local sample of stars is { on the whole quite} smooth.
Clearly, compared to the GCS sample, the LSS-GAC sample presented here is much less affected by the {bulk motions}, and thus should enable us to obtain a more robust estimate of $V_\odot$ than possible with the GSC sample.

\subsection{Determining $V_{\odot}$ by fitting the velocity distribution of the LSS-GAC sample}
To determine $V_{\odot}$, the model azimuthal velocity distribution is compared  to the observed one of the LSS-GAC sample.
The model azimuthal velocity distribution is generated  using the analytic formula derived by SB12.
As argued by SB12, this analytic formula provides excellent fit to the azimuthal velocity distribution yielded by rigorous torus-based dynamics modelling (Binney \& McMillan 2011) and reproduces the observed distribution of the GCS local sample of stars, which have accurate space velocity measurements.
We refer the reader to SB12 for a detailed discussion about this formula.
From the formula, the distribution of $V_{\phi}$ at a given Galactic position ($R$, $z$) is given by,
\begin{equation}
\begin{split}
n(V_{\phi}|R, z) = \mathcal{ N}\,e ^{-(R_{\rm g}-R_{0})/R_{\rm d}}\frac{2\pi R_{\rm g}K}{\sigma}\\
     \times\,{\rm exp}(-\frac{\Delta \Phi_{\rm ad}}{\sigma^{2}})\,f(z, R_{\rm g} - R)\,,
\end{split}
\end{equation}
where $\mathcal{ N}$ is a normalizing factor. $R_{\rm g}$  is the guiding-center radius given by, $R_{\rm g} = RV_{\phi}/V_{\rm c}$ (where $V_{\rm c}$ is the circular velocity as a function of $R_{\rm g}$, namely, the rotation curve. In this work we adopt a flat rotation curve of constant value 220 km s$^{-1}$).
$R_{0} = 8$ kpc is the Galactocentric distance of the Sun, $R_{\rm d}$ the disc scale length, $K$ a factor that can be numerically determined [cf. Eq (12) of SB12]. 
The term $\sigma$, being a function of $R_{\rm g}$, is given by
\begin{equation}
\sigma(R_{\rm g})\,=\,\sigma_{0}e^{-(R-R_{\rm g})/R_{\sigma}}\,,
\end{equation}
where $\sigma_{0}$ is the local velocity dispersion, $R_{\sigma}$ the scale length on which the velocity dispersion varies. 
${\Delta \Phi_{\rm ad}}$ is the so-called ``adiabatic potential'' that includes the effects of radial motion induced by adiabatic invariance of vertical motion [cf. Eq (33) of SB12 for detail].
$f(z, R_{\rm g} - R)$ is the $z$ factor (cf. Section 3.1 in SB12), which introduces other two parameters, $\alpha$\footnote{SB12 assume that the vertical force $K_{z}$ is proportional to $z^{\alpha-1}$. 
Most of the LSS-GAC sample stars are near the Galactic midplane and we adopt $\alpha = 1.5$ in the current work.} and the disc scale length $h_{\rm z}$.

We fit the $V_{\phi}$ distribution of our LSS-GAC sample by two components, a cool component representing the thin disc and another hot component representing the thick disc. 
The distribution of each of the two components are described by the above formula.
The parameters adopted for the two components are listed in Table 3.
To fit the data, each component is allowed to have two free parameters: the normalizing constant $\mathcal{N}$ and the local velocity dispersion $\sigma_{0}$. 
The combined distribution of $V_{\phi}$ of the two components is then calculated for a typical position ($R = 8.31$ kpc, $z = 0.04$ kpc) of our sample stars. 
The distribution is then shifted by an offset, which gives the value of $V_\odot$, in order to fit the observed distribution of the LSS-GAC sample.
Fig. 7 shows the best fit performed for the velocity range, $130 < V_{\phi} < 275$ km\,s$^{-1}$, on a linear and logarithmic scale.
The two component fit reproduces the observed distribution remarkably well.
For the whole sample, the best fit yields $V_{\odot} = 9.75 \pm 0.19$\,km\,s$^{-1}$.
The local velocity dispersions of the cool and hot components are 21.61 and 35.47 km\,s$^{-1}$, respectively.
The analysis is repeated {for the three} sub-samples, consisting respectively of stars of spectral type F, G and K, {selected  based on the values of effective temperature as described in Section 3.}

The typical position $(R, z)$, the number of stars $N$, as well as the value of $V_\odot$ deduced from the individual sub-samples, are listed in Table 4, along those for the whole sample.
Table 4 shows that values of $V_{\odot}$ yielded by the individual sub-samples are consistent with each other suggesting the robustness of the results.
The analyses also show that the local velocity dispersions of F-type stars are smaller than those of G- and K-type.
The result can be easily explained by the fact that stars of F-type are younger on average compared to those of G- and K-type, thus have smaller velocity dispersions.
The same effect also explains why the value of $V_{\odot}$ determined from F-type are slightly smaller than that deduced from G- or K-type stars.

\begin{figure}
\begin{center}
\includegraphics[scale=0.6,angle=0]{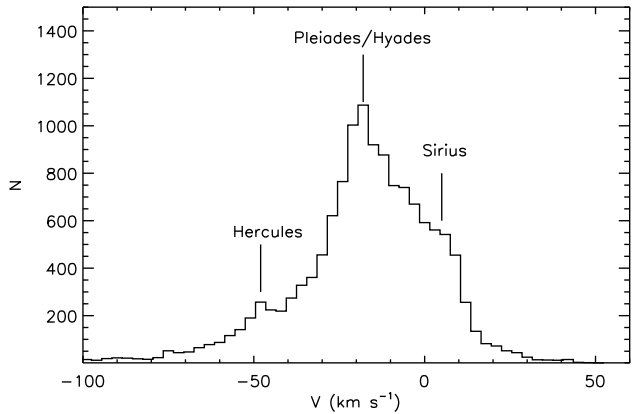}
\caption{Distribution of the $V$ velocity components of the GCS sample of local stars. 
                   Four well-known moving groups are clearly visible.}
\end{center}
\end{figure}

\begin{figure*}
\begin{center}
\includegraphics[scale=0.55,angle=0]{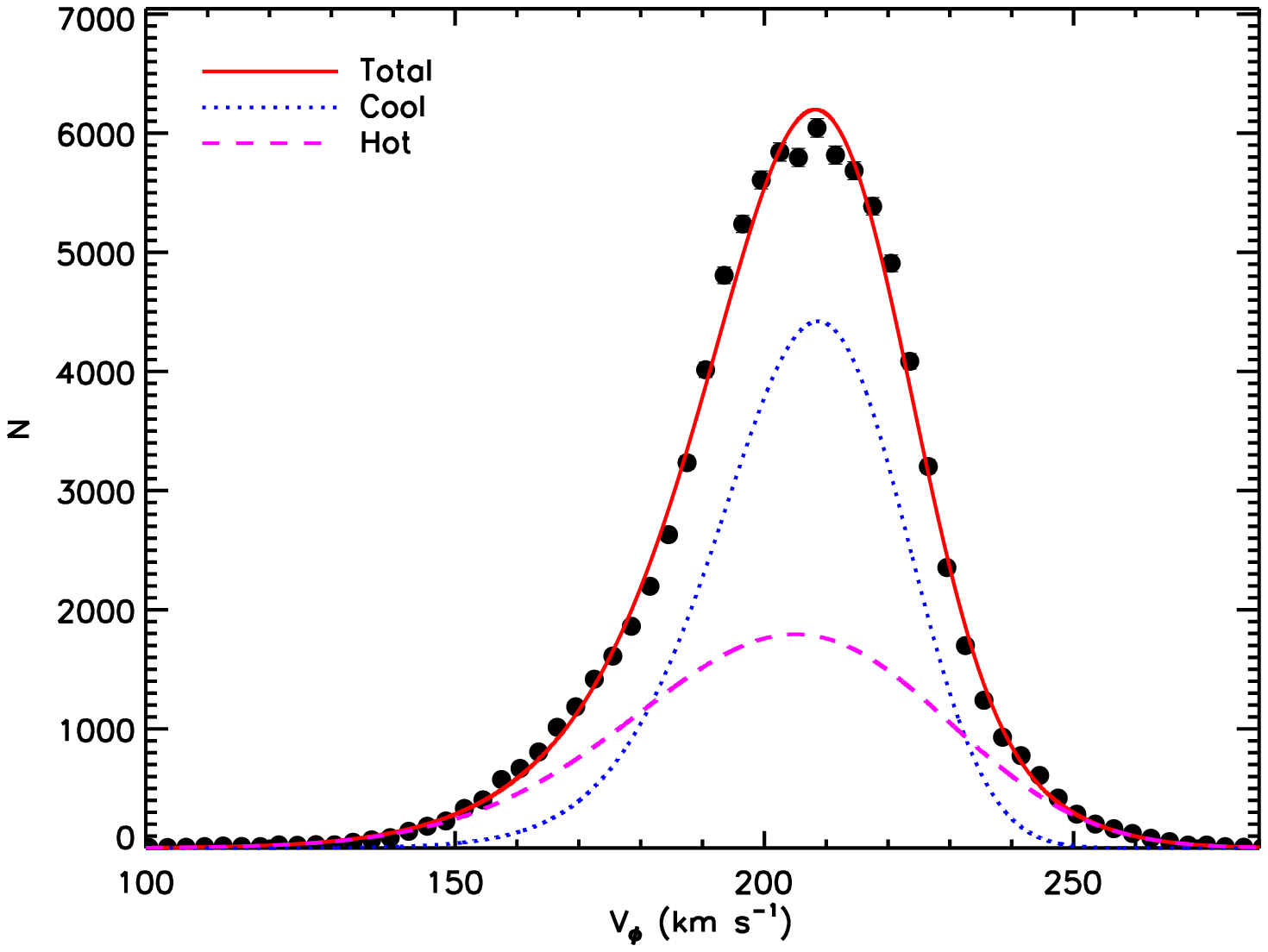}
\includegraphics[scale=0.55,angle=0]{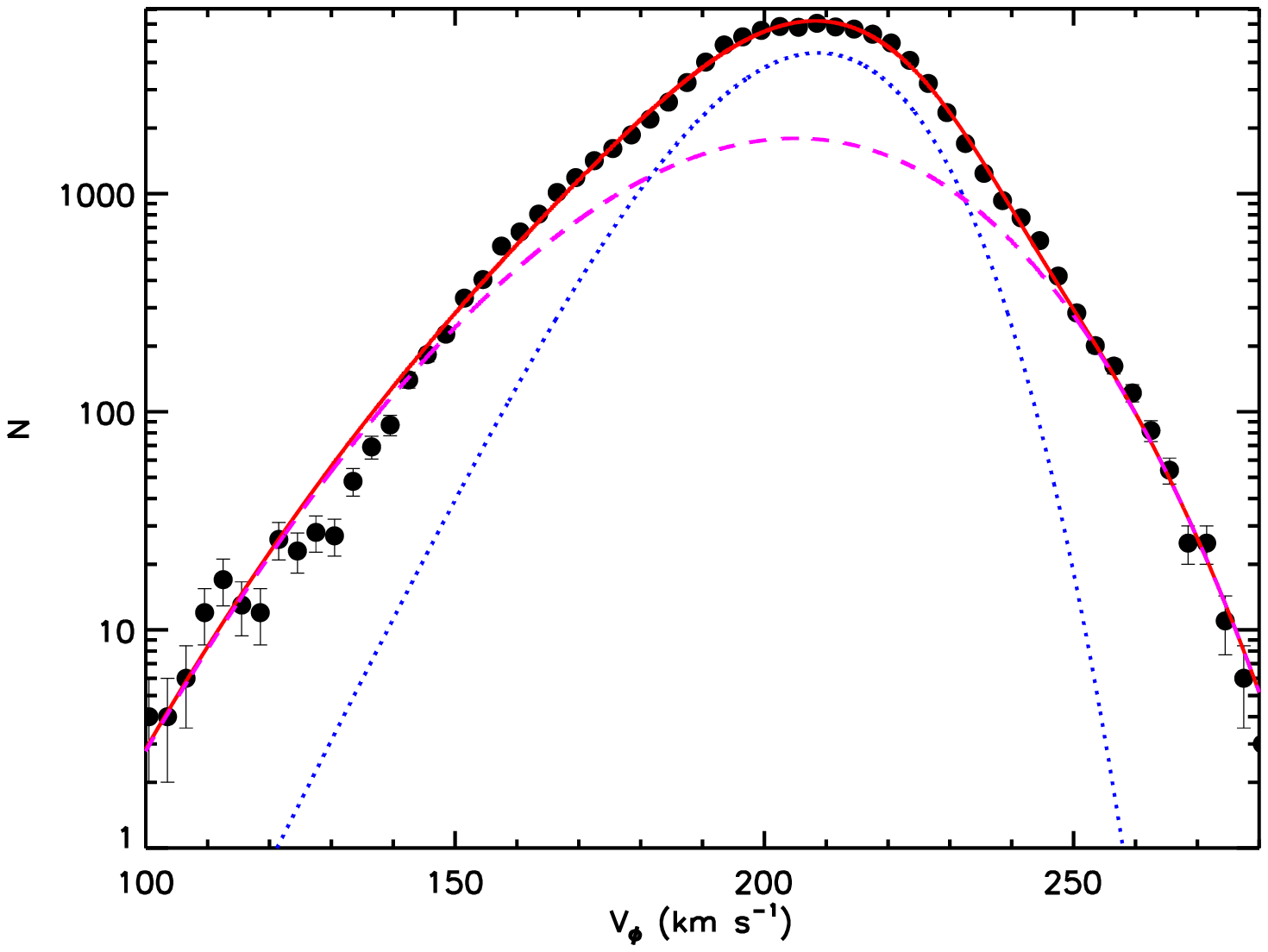}
\caption{The $V_{\phi}$ velocity distribution of the LSS-GAC sample of FGK main-sequence stars are fitted by the combination of two components: a cool component for the thin disc and a hot component  for the thick disc. 
                   Each component, characterized by distribution function Eq.\,(2),  has two free parameters: a normalization constant $\mathcal{N}$ and the local velocity dispersion $\sigma_{0}$.
                   To fit the data, the model distribution is shifted by a constant offset, which represents the peculiar velocity of the Sun, i.e.  $V_{\odot}$.
                   The left panel shows the model distributions and data points on a the linear scale, while in the right panel show the same distributions and data on a logarithmic scale, which reveals better the wings of the distributions.}
\end{center}
\end{figure*}

\begin{table}
 \begin{center} 
  \caption{Parameters adopted for the two components employed to fit the velocity distribution.}
  \begin{tabular}{cccc}
  \hline
   Components &                           $R_{\rm d}$ (kpc) & $h_{\rm z}$ (kpc) &$R_{\sigma}$ (kpc)\\
   \hline
   Cool  &2.4&0.2&7.5\\ 
   Hot&2.5&1.0&15.0\\        
 \hline
 \end{tabular}
\end{center}
\end{table}

\begin{table*}
 \center
  \caption{The solar motion in the Galactic rotation direction with respect to the LSR $V_{\odot}$ determined by VDF.}
  \begin{tabular}{lccccccc}
  \hline
   Sample &  $T_{\rm eff}$&  $V_{\odot}$&$R$&$Z$&$\sigma_{0}^{\rm cool}$&$\sigma_{0}^{\rm hot}$& $N$  \\
                  &   (K)               & (km s$^{-1}$)   & (kpc)   & (kpc) &(km s$^{-1}$)&(km s$^{-1}$)&      \\
   \hline
   Whole   &  $4200\,\le T_{\rm eff}\le\,6800$ & $9.75\pm0.19$ & $8.31$ & $0.03$&21.61&35.47 & 94332\\ 
   F type&  $6000\,\le T_{\rm eff}\le\,6800$ & $9.00\pm0.29$ & $8.38$ & $0.04$ & 19.02&26.85 &24220\\ 
  G type&  $5300\,\le T_{\rm eff}<\,6000$ & $10.03\pm0.21$ & $8.33$ & $0.05$ & 23.21&38.90&39100\\  
  K type&   $4200\,\le T_{\rm eff}<\,5300$ & $10.20\pm0.23$ & $8.24$ & $0.01$ & 22.27&38.29&31012\\         
\hline
\end{tabular}
\end{table*}

\section{Determining the solar motion by CTDS}
\subsection{Orbital eccentricities}
In addition to the technique of VDF, another method to determine the LSR is using only stars from the cold populations, thin disc that suffer from  negligible asymmetric drifts.
The method is simple and straightforward. 
Co\c skuno\u glu   et al. (2011) analyze the populations of a RAVE sample from which they select a sub-sample of thin disc stars based on their { Galactic space} velocity distribution, as proposed by Bensby et al. (2003, 2005), to derive the LSR. 
The result again confirms that previous estimates of $V_{\odot}$ from {Str\"omberg's equation} are grossly  underestimated by 6--7 km s$^{-1}$.
However, the {Galactic space} velocity distribution is not an efficient way to disentangle thin and thick disc stars, as shown by Binney (2010).
As a result, the sample of thin disc stars selected by Co\c skuno\u glu   et al. (2011) may be biased.
Here we use the orbital eccentricity [defined by Eq.\,(2)] as a natural indicator to define and select {cold populations of thin disc stars}.
Stars of low orbital eccentricities move in nearly circular orbits with negligible asymmetric drifts, whereas the orbits of those of stars with large eccentricities deviate from circular ones and thus possess significant asymmetric drifts.
In addition, by applying a stringent eccentricity cut, stars with significant bulk motions can also be screened out (Francis \& Anderson 2009).
By applying an eccentricity cut to the current sample, we expect to determine the LSR straightforwardly and robustly.
For this purpose, we use an eccentricity cut of 0.13, the peak value of thin disc stars (Lee et al. 2011). 
As shown by Bobylev and Bajkova (2014), the orbital eccentricities of young O to B2.5 stars,  used by them to derive the LSR range from 0 to 0.13.
The mean orbital eccentricity of Galactic open clusters is also found to be around 0.13 (Piskunov et al. 2006).
All those findings suggest that 0.13 is a reasonable eccentricity cut for the purpose of selecting thin disc stars.
{Furthermore}, we will show in Section 7 that the asymmetric drifts of {cold populations} selected using this orbital eccentricity cut are negligible.

The eccentricity of a star is defined by,
\begin{equation}
e = \frac{R_{\rm apo}-R_{\rm peri}}{R_{\rm apo}+R_{\rm peri}}\,,
\end{equation}
where $R_{\rm apo}$ ($R_{\rm peri}$) is the maximum (minimum) Galactocentric distance reached by the star in its orbit.
To calculate the eccentricities of stars of our FGK main-sequence sample, we integrate their orbits assuming a Galactic potential given by Gardner \& Flynn (2010).
In their model, the Milky Way consists of a Miyamoto \& Nagai (1975) disc, a Plummer (1911) bulge and inner core,  and a spherical logarithmic halo.
The characteristic parameters can be found in their Table 1.

\subsection{Impacts of the initial LSR }

\begin{table*}
 \begin{center} 
  \caption{Impacts of the initial LSR.}
  \begin{tabular}{cccccccc}
  \hline
   Sources &\multicolumn{3}{c}{Initial}  &\multicolumn{3}{c}{Final}  &$N$\\
   &$U_{\odot}$&$V_{\odot}$&$W_{\odot}$&$U_{\odot}$&$V_{\odot}$&$W_{\odot}$&\\
   &(km s$^{-1}$)&(km s$^{-1}$)&(km s$^{-1}$)&(km s$^{-1}$)&(km s$^{-1}$)&(km s$^{-1}$)&\\
   \hline
   Dehen \& Binney (1998)  &10.00&5.25&7.17&6.94&10.35&4.94&4\\
   Reid et al. (2009)& 9.00&20.00&10.00&6.98&10.38&4.95&5\\
   Co\c skuno\u glu et al. (2011) &8.50&13.38&6.49&6.99&10.39&4.95&4\\       
 \hline
 \end{tabular}
\end{center}
\end{table*}

To calculate orbital eccentricities, one however needs to assume a LSR.
In our approach, we first calculate orbital eccentricities with an initial guess of LSR and then iterate the entire LSR determination process until all components of the solar motion converge  to a degree less than the fitting errors (cf. Section 6.3).
In this approach, it is thus of utter importance to check and ensure that the final values of LSR thus derived are not sensitive to the initial values of { the LSR assumed}.
We have thus carried out extensive test of the impacts of initial LSR on the final results.
As shown in Table 5, we have used three vastly different sets of estimates of { the LSR} from the literature as our initial guess.
As pointed out earlier, the $U_{\odot}$ and $W_{\odot}$ components of the solar motion are hardly affected by the asymmetric drifts and straightforward to estimate. 
It is thus not surprising that their final derived values for the three completely different sets of initial guess do not differ much.
{It is remarkable that},  for the $V_{\odot}$ component which is known to suffer from significant asymmetric drift effects, the final estimates that result from the three completely different sets of initial guess that span a wide range of value from $\sim$ 5 to 20 km s$^{-1}$ all converge within one per cent!
In determining the LSR, we first calculate the orbital eccentricities of sample stars for each assumed set of initial values of  LSR and then apply orbital eccentricity cut to select {a sample of cold populations.}
The three components of the solar motions with respect to the LSR are then determined by the method described in the next subsection using the selected {sub-sample of cold populations.}
As shown in Table\,5, the final values of LSR converge after 4--5 iterations (the numbers of which are listed in the last column of the Table\,5).
The fact that the final values of  LSR are insensitive to the initial guess can be understood physically.
Stars belonging to the thick disc or halo population generally lag behind the rotation of thin disc stars by more than 45 km s$^{-1}$, {a lag that is more than two times the largest value} of the assumed initial ones of $V_{\odot}$. This implies that the orbital eccentricities of most thick disc or halo stars, calculated for the assumed initial LSR, remain {too large to pass} the orbital eccentricity cut, even for the case of { initial guess of} the largest value of $V_{\odot}$.
Meanwhile, most stars of our sample are near the Galactic midplane. 
The sample is thus dominated by {cold populations of thin disc stars.
Changes in the calculated eccentricities as induced by the different sets of initial LSR for most of those stars are quite small, and thus unlikely to significantly affect our selection of cold populations of thin disc stars using an eccentricity cut}.
At the beginning of the iteration process, there might be some contaminations  from those hot thick disc or halo stars. 
However, as the iteration progresses and the resultant {LSRs converge to} true values, the amount of contaminations {decreases}. 
It is thus quite natural that after several iterations, all results converge, even for quite different initial guess of LSR. 

The final distribution of orbital eccentricities of our FGK main-sequence sample is shown in Fig.\,8 (calculated as an example with the final LSR derived assuming the  initial LSR of DB98).
The sample consists of 60,595 thin disc stars selected by applying an eccentricity cut of less than 0.13.
The sample is used to derive the LSR in next subsection.

\begin{table*}
 \center
  \caption{Solar motion with respect to the LSR determined by the method of CTDS.}
  \begin{tabular}{lccccc}
  \hline
   Sample &  $T_{\rm eff}$& $U_{\odot}$            & $V_{\odot}$           & $W_{\odot}$       & $N$  \\
                  &   (K)               & (km s$^{-1}$)   & (km s$^{-1}$)   & (km s$^{-1}$) &      \\
   \hline
   Whole   &  $4200\,\le T_{\rm eff}\le\,6800$ & $6.94\pm0.25$ & $10.35\pm0.15$ & $4.94\pm0.09$ & 60595\\ 
   F type&  $6000\,\le T_{\rm eff}\le\,6800$ & $7.11\pm0.32$ & $10.06\pm0.17$ & $5.22\pm0.10$ & 17206\\ 
  G type&  $5300\,\le T_{\rm eff}<\,6000$ & $6.56\pm0.31$ & $10.27\pm0.18$ & $4.83\pm0.12$ & 23722\\  
  K type&   $4200\,\le T_{\rm eff}<\,5300$ & $7.27\pm0.31$ & $10.66\pm0.18$ & $4.78\pm0.11$ & 19667\\         
\hline
\end{tabular}
\end{table*}

\begin{figure}
\begin{center}
\includegraphics[scale=0.45,angle=0]{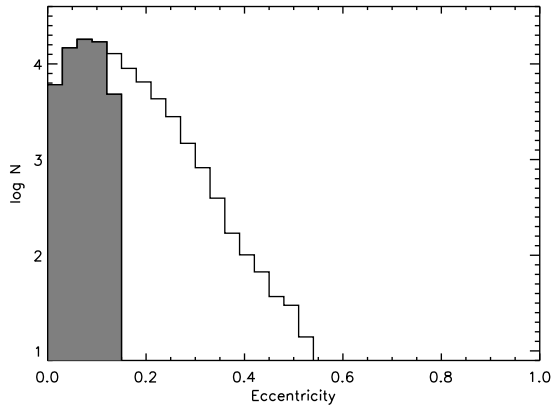}
\caption{Distribution of orbital eccentricities of sample. 
                   A total of sample 60,595 stars have eccentricities less than 0.13 and they comprise the subsample of thin disc stars used to calculate the LSR.}
\end{center}
\end{figure}

\begin{figure}
\begin{center}
\includegraphics[scale=0.45,angle=0]{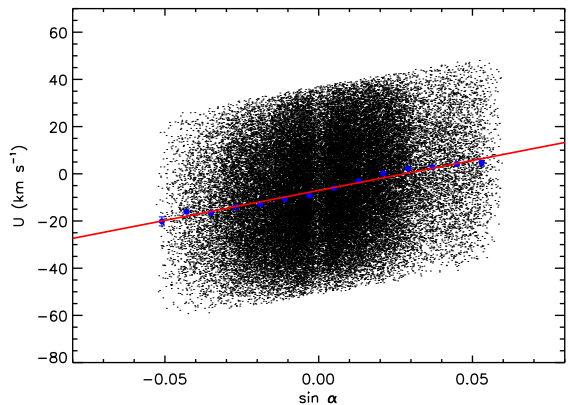}
\caption{$U$-component velocities are plotted against ${\rm sin}\,\alpha$  for the LSS-GAC thin disc sample of 60,595 FGK dwarfs. 
                    The blue dots represent the mean velocities of the individual bins of ${\rm sin}\,\alpha$ and the red line is a linear fit to the dots. }
\end{center}
\end{figure}

\subsection{Corrections for differential Galactic rotation and determinations of the solar motions }
\begin{figure*}
\begin{center}
\includegraphics[scale=0.6,angle=0]{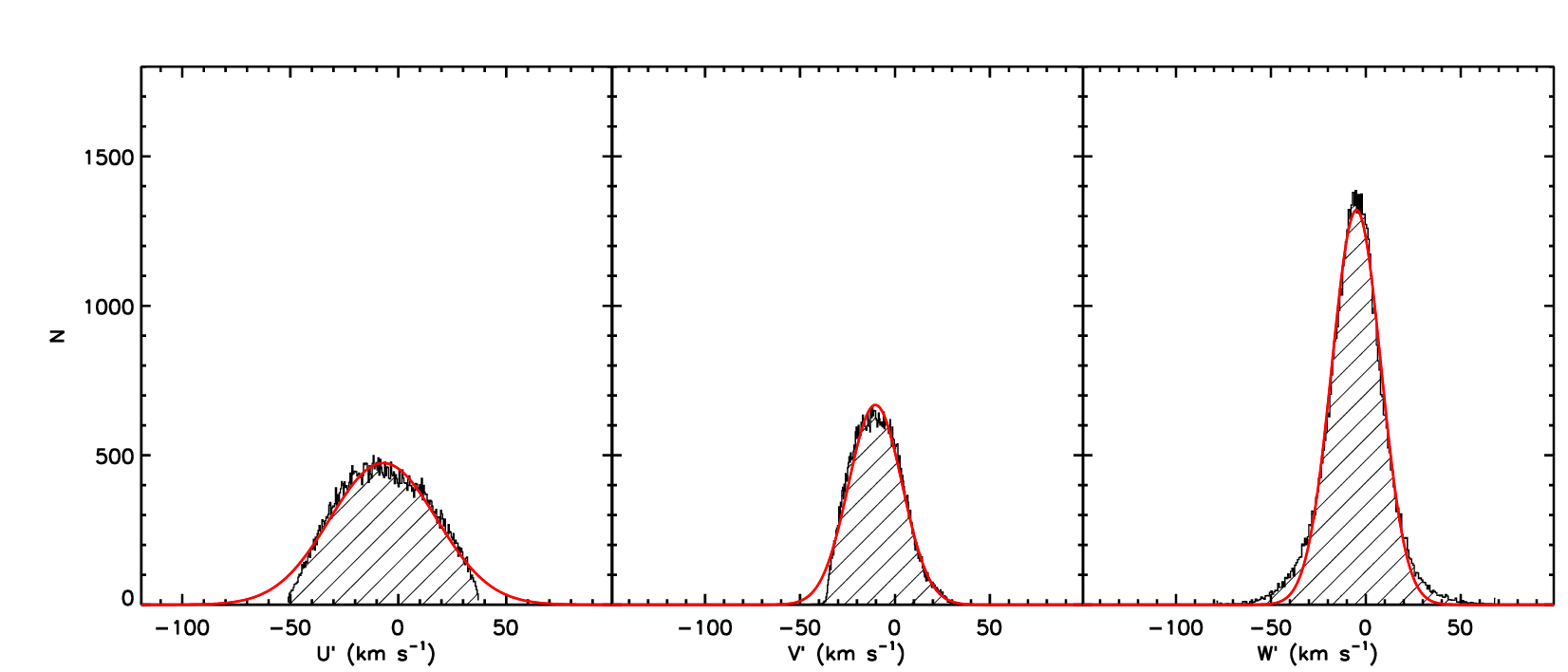}
\caption{Distributions of the three velocity components of a sample of 60,595 cold thin disk stars, after correcting for { differential Galactic rotation in radial and azimuthal direction ($U'$ and $V'$)}. Also overplotted are Gaussian fits to the distributions.
                   Note that the distribution of { $U'$} is truncated due to the lack of stars of high eccentricities in the current sample as a result of the orbital eccentricity cut.
                   Estimate of the mean of the distribution, which gives $-U_\odot$, is however not affected by the truncation.}
\end{center}
\end{figure*}

Considering { differential Galactic rotation}, the Galactic space velocity components of a star measured by an observer at the Solar position can be expressed as {follows}:
\begin{equation}
U = V_{\rm c}^{\rm s}\,{\rm sin}\,\alpha - U_{\odot} + \epsilon_{U}\,,
\end{equation}
\begin{equation}
V = (V_{\rm c} - V_{\rm c}^{\rm s}{\rm cos}\,\alpha) -  V_{\odot} + \epsilon_{V}\,,
\end{equation}
\begin{equation}
W = - W_{\odot} + \epsilon_{W}\,,
\end{equation}
where $\alpha$ is the angle between the star and the Sun with respect to the Galactic centre,
$V_{\rm c}$ is the circular speed of the Sun,
$V_{\rm c}^{\rm s}$ represents the { mean azimuthal speed of the selected stellar population} of concern here, and $\epsilon_{U}$, $\epsilon_{V}$, $\epsilon_{W}$ are respectively the noise values in the three velocity component $U$, $V$ and $W$, resulting from the combined effects of the velocity measurement uncertainties $\delta_{U}$, $\delta_{V}$, $\delta_{W}$ and the (intrinsic) velocity dispersions $\sigma_{U}$, $\sigma_{V}$, $\sigma_{W}$.

In order to determine the LSR, two approaches are generally used to account for the effects of differential Galactic rotation in Eqs.\,(5) and (6).
The first approach is to fit the observed values of  $U$ and $V$ as a function of $\alpha$ { (Sch\"onrich et al. 2012)}.
Alternatively, one can take a flat rotation curve with a constant $V_{\rm c} = 220$\,km s$^{-1}$, and then subtract the terms of differential  Galactic rotation from the observed values of  $U$ and $V$ in Eqs.\,(5) and (6).     
Here we should point out that the 600\,pc distance cut discussed above is needed for both the approaches to work, since the cut effectively reduces the effects of large-scale streaming motions, such as the radial velocity gradient and pattern in an extended disc (Siebert et al. 2011, Williams et al. 2013), induced by, for example, non-axisymmetric structures (e.g. central bar, spiral arms). 
The large number of stars peaking at 600\,pc in our sample (Fig.\,3) also ensures that a sufficient number of stars is available for  robust LSR determination.

 The magnitude of streaming motions in $U$, represented by the term  $V_{\rm c}^{\rm s}$\,${\rm sin}\,\alpha$ in Eq.\,(5),  is of the same order of magnitude of the velocity { noise}, on the level of dozens of km\,s$^{-1}$ for stars of the current sample.
However, the range of variations of the similar term in $V$, i.e. $V_{\rm c}^{\rm s}$\,${\rm cos}\,\alpha$, is less than 1 km s$^{-1}$, and thus will be lost in the velocity { noise}.
Thus the first approach can only be used to determine $U_{\odot}$.
Fig.\,9 plots the velocity component $U$ against sin\,$\alpha$ for 60,595 thin disc stars, selected from our LSS-GAC sample of FGK dwarfs after imposing an eccentricity cut of 0.13. 
The DB98 {initial LSR} is assumed. 
A linear fit to the data yields ($U_{\odot}$,\,$V_{\rm c}^{\rm s}$) = ($7.03\pm0.35$, $254.00\pm10.54$) km s$^{-1}$.
The value of $V_{\rm c}^{\rm s}$ is not constrained by the current sample due to the relatively  small range of $\sin \alpha$  covered by the sample.

For the second approach, after correcting for { differential Galactic rotation}, only the solar peculiar velocities and the velocity { noise} remain. 
The noise value of each velocity component is assumed to obey a Gaussian distribution (Bensby et al. 2003, 2005) of dispersions $\zeta_{U,V,W}=\sqrt{\delta_{U,V,W}^{2}+\sigma_{U,V,W}^{2}}$.
Thus the space velocities after corrections should also follow Gaussian distributions, { $U' = U - V_{c}\sin\alpha \sim$\,\textbf{G}\,($-{U_{\odot}},\,\zeta_{U}^{2}$), $V' = V- (V_{c} - V_{c}\cos\alpha) \sim$\,\textbf{G}\,($-{V_{\odot}},\,\zeta_{V}^{2}$) and $W' = W \sim $\,\textbf{G}\,($-{W_{\odot}},\,\zeta_{W}^{2}$)}. 
Distributions of the three velocity components, after correcting for { differential Galactic rotation}, for the current sample of 60,595 thin disc stars are presented in Fig.\,10, for the case of { the DB98} initial LSR.
Also overplotted in the Figure are Gaussian fits to the distributions.
{ The corrected velocity distributions, $U'$ and $V'$, are well described by a Gaussian function. }
The velocity distribution in radial component is however truncated as a result of the orbital eccentricity cut applied to the sample.
{The estimate} of the mean of the distribution, which {yields} $-U_\odot$, is however not affected by the truncation. 
The fits presented in Fig.\,10 yield
{($U_{\odot}$, $V_{\odot}$, $W_{\odot}$) = ($6.94\pm0.25$, $10.35\pm0.15$, $4.94\pm0.09$).}
The value of $U_{\odot}$ derived from the Gaussian fit is consistent with that yielded by the first approach, i.e. by fitting { differential Galactic rotation}, as  given above.

\begin{table}
 \center
  \caption{Final adopted solar motions with respect to the LSR.}
\begin{threeparttable}
  \begin{tabular}{lccc}
  \hline
   Methods & $U_{\odot}$            & $V_{\odot}$           & $W_{\odot}$   \\
                                 & (km s$^{-1}$)   & (km s$^{-1}$)   & (km s$^{-1}$)      \\
   \hline
   VDF \tnote             &   -- &$9.75\pm0.19$ & -- \\
   CTDS\tnote{a,*}  &   $6.97\pm0.25$ & $10.37\pm0.15$ & $4.95\pm0.09$ \\  
   CTDS\tnote{b,*}    &   $7.08\pm0.34$ & -- & -- \\  
  Adopted     &   $7.01\pm0.20$ & $10.13\pm0.12$ & $4.95\pm0.09$ \\         
\hline
\end{tabular}
\begin{tablenotes}\small
\item[a] { After corrected differential Galactic rotation assuming a flat rotation curve of constant $V_{c}=220$\,km\,s$^{-1}$.} 
\item[b] {After corrected differential Galactic rotation by fitting Eq.\,(5) (see also Fig.\,9)}.
\item[*] The value of each component is the mean of results deduced from the three sets of initial LSR listed in Table 5.
 \end{tablenotes}
\end{threeparttable}
\end{table}

\begin{figure*}
\begin{center}
\includegraphics[scale=0.6,angle=0]{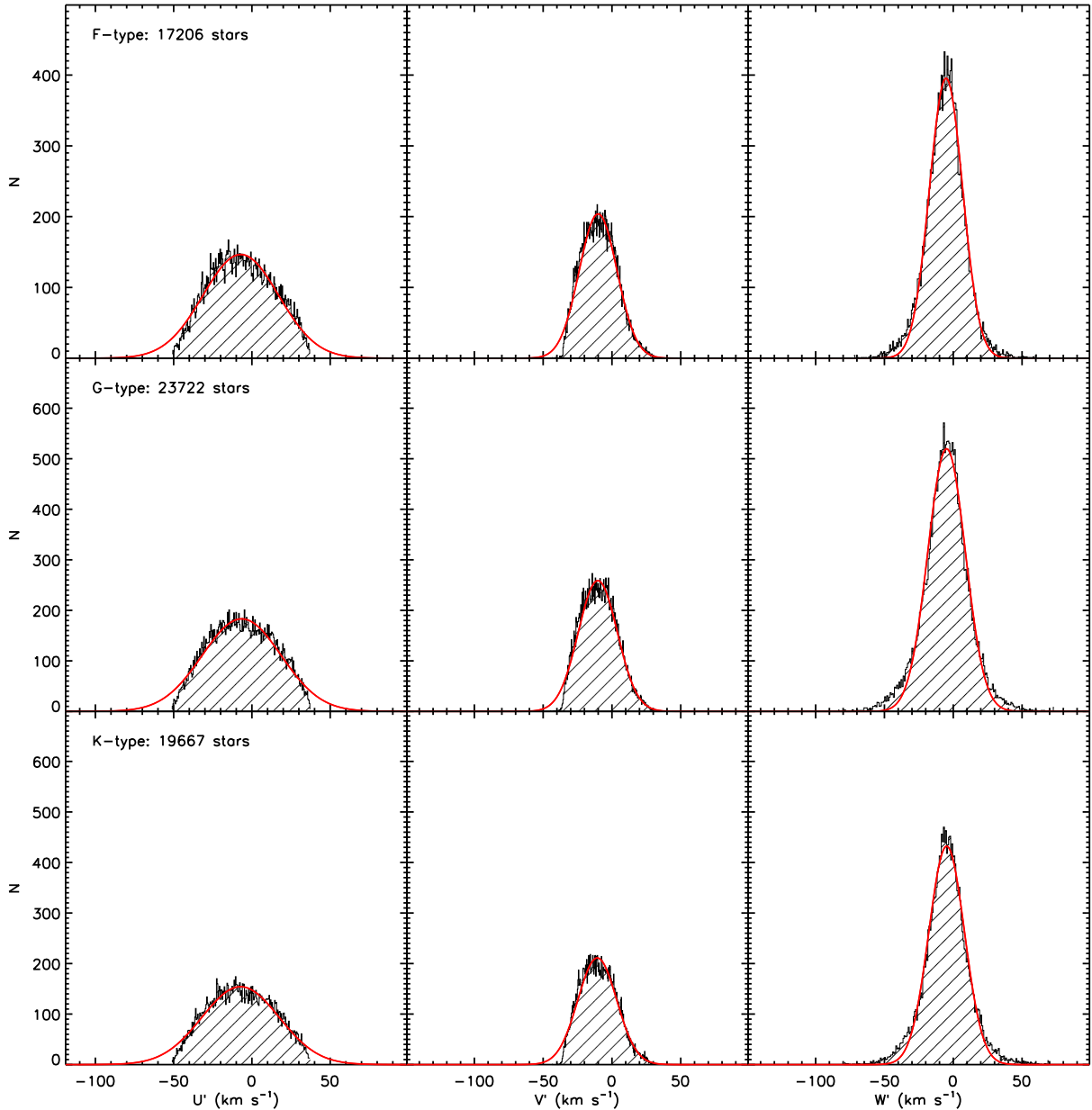}
\caption{Same as Fig.\,10 but for sub-samples of different spectral types. The numbers of stars of the individual spectral types are marked in the left panels of the Figure. }
\end{center}
\end{figure*} 

\section{DISCUSSION}
The value of $V_{\odot}$ obtained in the current work by { the VDF method} is 1--2 km s$^{-1}$  smaller than those of Binney (2010) and Sch\"onrich et al. (2010).
The small differences might be caused by bulk motions that affect the samples used in these work, as discussed earlier.
 As Fig.\,6  shows, the presence of the Pleiades and/or Hyades streams moves the peak of $V$ distribution of the GCS local sample toward negative values,  leading to their estimates of $V_\odot$ being1--2 km s$^{-1}$ larger than obtained with the LSS-GAC sample, which shows a much smoother azimuthal velocity distribution and is less affected by the bulk motions of stellar streams.

For the CTDS method, we have determined the LSR based on two assumptions.
Firstly, we assume that {cold populations of thin disc stars} selected by applying an orbital eccentricity cut suffer from negligible asymmetric drift effects.
The asymmetric drifts of { cold populations of  thin disc stars selected from} the orbital eccentricity cut can be estimated approximately using  Eq.\, (4.228) of Binney \& Tremaine (2008) from the velocity dispersions ($\sigma_{U,V,W}$), velocity ellipsoid, stellar density ($\nu$) and circular speed ($V_{\rm c}$).
Assume that the tilt angle is toward the Galactic center (Binney et al. 2014; B\"udenbender et al. 2014) and  both $\nu$ and $\sigma_{U}$  are exponential functions of $R$,
\begin{equation}
\nu \propto {{\exp}}(-\frac{R}{R_{d}})\,,
\end{equation}
\begin{equation}
\sigma_{U} \propto {{\exp}}(-\frac{R}{R_{\sigma}})\,,
\end{equation}
where $R_{d}$ and $R_{\sigma}$ are the scale lengths of $\nu$ and $\sigma_{U}$, respectively.
Under those assumptions, Eq.\, (4.228) of Binney \& Tremaine (2008) becomes,
\begin{equation}
V_{a} \simeq\frac{1}{2V_{c}}\left[\sigma_{V}^{2} + \sigma_{W}^{2} + R\sigma_{U}^{2}(\frac{1}{R_{d}} + \frac{2}{R_{\sigma}} - \frac{2}{R})\right]\,,
\end{equation}
where the intrinsic velocity dispersions\footnote{The  intrinsic velocity dispersions are estimated by subtracting the space velocity errors using the sample of 60,595 thin disc stars described in Section 6.2.} of the selected thin disc stars are $\sigma_{U} = 18.98$, $\sigma_{V} = 11.05$  and $\sigma_{W} = 9.61$ km\,s$^{-1}$, respectively.
Adopting $R_{\sigma} = 13.70$ kpc from Sharma et al. (2014), $R_{d} = 2.5$ kpc, $V_{c} = 220$ km\,s$^{-1}$ and a typical position $R = 8.3$ kpc of the selected stars, we have $V_{a} = 2.56$ km\,s$^{-1}$.
If using instead a larger scale length $R_{d} = 3.7$ kpc from the recent work of Chang et al. (2011),  the estimated value of $V_{a}$ decrease to 1.68 km\,s$^{-1}$.


{The results show that our assumption that cold populations of thin disk stars selected with an orbital eccentricity cut of $e$\,$< 0.13$ have small asymmetric drifts is a reasonable one.
 This is corroborated by the close agreement of values of $V_\odot$ deduced from the two independent analyses, i.e. those based on the VDF and CTDS methods, respectively.
The effects of asymmetric drifts of the current sample of thin disc stars on $V_\odot$ estimation, if any, are likely to be quite small\footnote{{ Considering the fact that the value of $V_\odot$ deduced from the CTDS} {method is only 0.62 km\,s$^{-1}$ {\em larger} than that deduced from the VDF method  (which fully accounts for the effects of asymmetric drifts), the value of $V_{a}$ of the current sample of thin disc stars is likely to be smaller than 1 km\,s$^{-1}$.}}.}
As a further test of the assumption, we note that stars of different spectral types (populations) suffer from different levels of the asymmetric drift effects.
For sub-samples of dwarfs of F, G and K spectral type, the level of those effects can vary from a few to tens of km s$^{-1}$ (DB98, Sch\"onrich et al. 2010).
The fact that all the three of F, G and K sub-samples give consistent results suggest again the assumption is a reasonable one.



{ As described earlier, the second assumption underlying our analysis with the CTDS method is that the sample suffers from little systematic kinematic effects, induced by, for example, non-axisymmetric structures, accretion events or asymmetric mass infall.
As already pointed out in Section 5.1, the current sample does not seem to be affected significantly by any moving group.}
The validity of this assumption also seems to be supported by the consistent results yielded by the individual sub-samples of different spectral type, considering that those sub-samples probe different depths (because of their different intrinsic, absolute magnitudes)\footnote{ Here, we are not arguing that cold populations are less affected by those non-axisymmetric structures than any warm tracer.
We just point out, by good fortune or not, that the current sample of cold populations seems to happen to be hardly affected by those structures.}.
Fig.\,11 plots the results from the three sub-samples for the case of {the DB98} initial LSR.
The values of solar motion derived from the sub-samples are presented in Table\,6.
Fig.\,11 and Table\,6 show that for all the three velocity components of LSR, the values yielded by all the three sub-samples agree within the fitting errors.
As a by-product, the results also suggest that  the local disk seems to be radially and vertically stable and circularly symmetric.


We note all the errors quoted for the estimates of LSR presented here are purely formal.
There are potential systematic errors for both the methods of VDF and CTDS.
For the VDF method, potential biases may come from the underlying assumptions and approximations, as well as from the parameters adopted to construct the analytical azimuthal velocity distribution.
As for the CTDS method, we believe that the effects of asymmetric drift are minor, although effects on the level of $\sim 1$--2 km s$^{-1}$ could be present as estimated above, considering that the thin disc stars selected do not have exactly zero eccentricity, i.e. they do not move in perfect circular orbits.
Although there might be a variety of potential sources of systematic errors affecting the results from both methods, it is { reassuring} that both methods yield consistent values of  $V_{\odot}$ within 2$\sigma$.
Our final, adopted values of LSR are presented in Table 7, obtained by  taking the mean of individual estimates, weighted by the inverse square of the formal errors of individual estimates.

\section{CONCLUSIONS}
The traditional method based on { Str\"omberg's equation} fails to produce accurate estimates of the LSR, especially for the $V_{\odot}$ component,  due to the presence of kinematic and/or metallicity biases addressed recently by a number of investigators.
In this work, two alternative and independent methods have been used  to derive the LSR using a completely new sample of stars selected from the  LSS-GAC DR1.

The first method derives $V_{\odot}$ by fiting the observed azimuthal velocity distribution with {an analytical} distribution function which has been shown to reproduce well reults from rigorous torus-based dynamics modeling.
We obtain $V_{\odot} = 9.75\pm0.19$\,km\,s$^{-1}$.
The value is about 1--2 km\,s$^{-1}$ smaller than those derived by Binney (2010) and Sch\"onrich et al. (2010) by applying the same method to the GCS local sample.
We suggest that the small differences are caused by the presence of stellar streams that significantly plagues the GCS sample, but to a much less extent the LSS-GAC DR1 sample.

The second method derives the LSR directly using { cold populations of thin disc stars} as tracers.
The method relies on the definition of a proper sample of {cold populations of thin disc stars}.
In this paper, we propose that the orbital eccentricity serves as a natural quantity to select {cold populations of thin disc stars}. 
After applying an orbital eccentricity cut, two approaches are adopted to correct for the effects of differential Galactic rotation and determine  the LSR.
The first approach is applicable to the $U_\odot$ component only.
The analysis yields $U_{\odot}\,=\,7.03\pm0.35$ km s$^{-1}$. 
The second approach, under the assumption of a flat rotation curve and negligible asymmetric drifts,  is based on Gaussian fitting to the measured velocity distributions, after correcting for the effects of differential Galactic rotation.
Values of $U_{\odot}$ derived from both approaches are consistent with each other and in agreement with that deduced from maser  measurements (Bobylev \& Bajkova 2010, 2014). 
From the second approach, we find $V_{\odot}\,=\,10.35\pm0.15$ km s$^{-1}$, which is consistent within 2$\sigma$ with the value deduced by the method of VDF. 
Our newly derived value of $W_{\odot}\,=\,4.94\pm0.09$ km s$^{-1}$ is very close to the recent estimate of Sch\"onrich (2012).
Both for VDF and CTDS methods, the whole sample is further divided into sub-samples of F, G and K spectral types. 
Results derived from those sub-samples all yield consistent results, as listed in Table \,4 and \,6. 
 The results corroborate the robustness of the LSR determined by both methods.

Benefited from the LSS-GAC DR1 and the UCAC4  proper motion catalog, we have built the largest local sample of FGK main-sequence stars within 600 pc of the Sun with accurate distances and velocities.
By controlling the random errors discussed in Section 4 and correcting systematic errors both in radial velocities and in proper motions as discussed in Section 3.2, the sample has enabled us to determine the LSR with the smallest uncertainties hitherto (see Table 1).
Our final recommended values of LSR are:  ($U_{\odot}$, $V_{\odot}$, $W_{\odot}$) = ($7.01\pm0.20$, $10.13\pm0.12$, $4.95\pm0.09$) km s$^{-1}$ (see Table\,7).

 \section*{Acknowledgements} 
 It is a pleasure to thank Dr. J. Binney for a critical reading of the manuscript and constructive comments.
 This work is supported by National Key Basic Research Program of China 2014CB845700.  
 
 The Guoshoujing Telescope (the Large Sky Area Multi-Object Fiber Spectroscopic Telescope, LAMOST) is a National Major Scientific Project built by the Chinese Academy of Sciences. Funding for the project has been provided by the National Development and Reform Commission. LAMOST is operated and managed by the National Astronomical Observatories, Chinese Academy of Sciences.
 

\end{document}